\documentclass[conference]{IEEEtran}
\IEEEoverridecommandlockouts

\usepackage{cite}
\usepackage{amsmath,amssymb,amsfonts}

\usepackage{textcomp}
\usepackage{xcolor}

\def\BibTeX{{\rm B\kern-.05em{\sc i\kern-.025em b}\kern-.08em
    T\kern-.1667em\lower.7ex\hbox{E}\kern-.125emX}}

\usepackage[noend]{algpseudocode}
\usepackage{graphicx,subcaption}
\usepackage{lipsum}
\usepackage{algorithm,algcompatible,amsmath}
\usepackage{color}
\usepackage[font=small,labelfont=bf]{caption}

\algnewcommand\INPUT{\item[\textbf{Input:}]}%
\algnewcommand\OUTPUT{\item[\textbf{Output:}]}%
\algnewcommand\CHECK{\item[\textbf{Check:}]}%
\algnewcommand\COMPUTE{\item[\textbf{Compute:}]}%

\title{A biologically constrained encoding solution for long-term storage of images onto synthetic DNA}

\author{\IEEEauthorblockN{Dimopoulou Melpomeni}
\IEEEauthorblockA{\textit{Universit\'e C\^ote d'Azur} \\
\textit{I3S,CNRS,UMR 7271}\\
Sophia Antipolis, France \\
dimopoulou@i3s.unice.fr}
\and
\IEEEauthorblockN{Marc Antonini}
\IEEEauthorblockA{\textit{Universit\'e C\^ote d'Azur} \\
\textit{I3S,CNRS,UMR 7271}\\
Sophia Antipolis, France \\
am@i3s.unice.fr}
\and
\IEEEauthorblockN{Pascal Barbry}
\IEEEauthorblockA{\textit{Universit\'e C\^ote d'Azur} \\
\textit{IPMC, CNRS, UMR 7275}\\
Sophia Antipolis, France \\
barbry@impc.cnrs.fr}
\and
\IEEEauthorblockN{Raja Appuswamy}
\IEEEauthorblockA{\textit{EURECOM} \\
Sophia Antipolis, France \\
appuswam@eurecom.fr}
}

\begin{document}

\maketitle

\begin{abstract}
Living in the age of the digital media explosion, the amount of data that is being stored increases dramatically. However, even if existing storage systems suggest efficiency in capacity, they are lacking in durability. Hard disks, flash, tape or even optical storage have limited lifespan in the range of 5 to 20 years. Interestingly, recent studies have proven that it was possible to use synthetic DNA for the storage of digital data, introducing a strong candidate to achieve data longevity. The DNA's biological properties  allows the storage of a great amount of information into an extraordinary small volume while also promising efficient 
storage for centuries or even longer with no loss of information.  However, encoding digital data onto DNA is not obvious, because when decoding, we have to face the problem of sequencing noise robustness. Furthermore, synthesizing DNA is an expensive process and thus, controlling the compression ratio by optimizing the rate-distortion trade-off is an important challenge we have to deal with. \\
This work proposes a coding solution for the storage of digital images onto synthetic DNA. We developed a new encoding algorithm which generates a DNA code robust to biological errors coming from the synthesis and the sequencing processes. Furthermore, thanks to an optimized allocation process the solution is able to control the compression ratio and thus the length of the synthesized DNA strand. Results show an improvement in terms of coding potential compared to previous state-of-the-art works.
\end{abstract}


\begin{figure*}[htb]
\centering
    \includegraphics[width=0.9\textwidth]{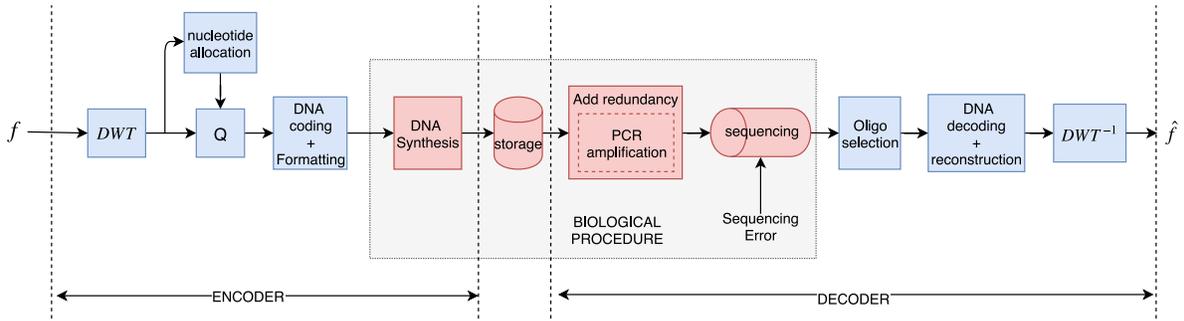}
\caption{the general encoding schema \vspace{-3mm}} 
\label{fig:general_schema}
\end{figure*}

\section{Introduction}
Storage of digital data is becoming challenging for humanity due to the relatively short life span of storage devices. At the same time, the digital universe (all digital data worldwide) is forecast to grow to over 160 zettabytes in 2025. A significant fraction of this data is called cold or infrequently accessed. Old photographs stored by users on Facebook is one such example of cold data; Facebook recently built an entire data center dedicated to storing such cold photographs. Unfortunately, all current storage media used for cold data storage (Hard Disk Drives or tape) suffer from two fundamental problems. First, the rate of improvement in storage density is at best 20\% per year, which substantially lags behind the 60\% rate of cold data growth. Second, current storage media have a limited lifetime of five (HDD) to twenty years (tape). As data is often stored for much longer duration (50 or more years) due to legal and regulatory compliance reasons, data must be migrated to new storage devices every few years, thus, increasing the price of data ownership. An alternative approach may stem from the use of DNA, the support of heredity in living organisms.  DNA possesses three key properties that make it relevant for archival storage. First, it is an extremely dense threedimensional storage medium that has the theoretical ability to store 455 Exabytes in 1 gram; in contrast, a 3.5” HDD can store 10TB and weighs 600 grams today. Second, DNA can last several centuries even in harsh storage environments; HDD and tape have life times of five and thirty years. Third, it is very easy, quick, and cheap to perform in-vitro replication of DNA; tape and HDD have bandwidth limitations that result in hours or days for copying large EB-sized archives.
DNA is a complex molecule corresponding to a succession of four types of nucleotides (nts), Adenine (A), Thymine (T), Guanine (G), Cytosine (C). It is this quaternary genetic code that inspired the idea of DNA data storage which suggests that any binary information can be encoded into a DNA sequence of A, T, C, G. 
The main challenge lies in the restrictions imposed by the biological procedures of DNA synthesis (writing) and sequencing (reading) which are involved in the encoding process and introduce significant errors in the encoded sequence while also being relatively costly (several dollars for writing and reading a small strand of nucleotides).

 Recent works tackle the problem of digital data storage onto DNA still leaving room for further improvements that could finally bring the idea of DNA data storage into practice. In \cite{b1} there has been a first attempt to store data into DNA while also providing a study of the main causes of biological error. In order to deal with errors previous works in \cite{b2} and \cite{b3} have suggested dividing the original file into overlapping segments so that each input bit is represented by multiple oligos. However, this procedure introduces extra redundancy and is poorly scalable. Other studies \cite{b4},\cite{b5} suggest the use of Reed-Solomon code in order to treat the erroneous sequences while in \cite{b6} a new robust method of encoding has been proposed to approach the Shannon capacity. Finally, latest works in \cite{b7} have introduced a clustering algorithm to provide a system of random-access DNA data storage. Nevertheless, all these approaches mainly try to convert a binary bit stream onto a DNA sequence without considering the original input data characteristics. In addition to this, as the DNA synthesis cost can be really high it is extremely important to take full advantage of the optimal compression that can be achieved before synthesizing the sequence into DNA. Although previous works have used compressed data such as images in a JPEG format the final encoding has been carried out on the compressed bit stream without interfering to the compression procedure.
 
 In this paper we make a very first step in introducing image compression techniques for long term image storage onto synthetic DNA. One of our main goals is to allow the reduction of the cost of DNA synthesis which nowadays can be very high for storage purposes. Unlike previous works that have been transcoding directly binary sequences onto DNA, our coding algorithm is applied on the quantized wavelet coefficients of an image (as shown in figure \ref{fig:general_schema}). To this end, the proposed solution is optimized thanks to a nucleotide allocation process across the different wavelet subbands by taking into account the input data characteristics. The desired compression rate can then be chosen, allowing to control the DNA synthesis cost. Furthermore, we have developed a new encoding algorithm which generates a DNA code robust to biological errors coming from the synthesis and the sequencing processes.
 In section 2 of this paper, we describe the general encoding process analyzing the biological restrictions and the creation of the codewords of the  coding dictionary that will be used. Furthermore, we present the exact formatting of our encoded data and define the procedures of DNA synthesis and storage. In section 3, we analyse data decoding method by explaining the process of DNA sequencing. Finally, in section 4 we demonstrate our results and in section 5 we conclude by proposing some interesting future steps of this work.

\section{The proposed encoding scheme}
\label{sec:enc}

\subsection{The general idea}
\label{ssec:general_idea}
The main goal of DNA data storage is the encoding of the input data  using a quaternary code composed by the alphabet \{A,T,C,G\}to be stored into DNA. The general idea of our proposed encoding process is depicted in figure \ref{fig:general_schema} and can be very roughly described by the following steps. Firstly, the input image has to be compressed using a lossy/near lossless image compression algorithm.  Here, we propose to use a discrete wavelet decomposition (DWT) to take advantage of the spatial redundancies of the image creating subbands which are being compressed using a uniform scalar quantizer $Q$. The quantization step size $q$ is selected according to an optimal nucleotide allocation algorithm that minimizes the distortion on the compressed image while constraining the length of the nucleotide strand generated by the encoder in each wavelet subband. After quantization, each subband is encoded into a sequence of A,T,C and G to be later synthesized into DNA. Error in DNA synthesis and yield of production can be  well controlled when the oligonucleotide size stays below 150 nts. This implies that the encoded sequences needs to be cut into smaller chunks before synthesis. Stability of oligos can be good for several months at -20\textdegree C. For longer storage (from years to centuries, and probably over), DNA must be encapsulated in DNAshell  (in our experiment, this operation was done by the company Imagene, Evry, France), a storage that protects it from contacts with oxygen and water. DNA sequencing is the process of retrieving the stored information by reading the content of the stored oligos. Unfortunately this procedure is very error prone causing errors like substitutions, insertions or deletions of nucleotides. In order to deal with such errors, before sequencing, the stored data is cloned into many copies using a biological process called Polymerace Chain Reaction (PCR) amplification. In addition to this, during the sequencing, next generation sequencers (NGS) like Illumina use the method of bridge amplification (BA) for reading the oligos. As exlained in \cite{b8}, BA is a process similar to PCR, which allows nucleotide recognition while producing many copies of each oligo introducing extra redundancy that is necessary for the reduction of the sequencing error. As a result the data provided by the sequencer is multiple copies of each oligo that may contain errors. This implies the need of choosing the good oligo copies before we reconstruct the initial sequence concatenating the selected oligos and decode them  to get back the stored data. 

\subsection{Biological Restrictions}
\label{ssec:restrictions}
As described in \ref{ssec:general_idea} the biological procedure of DNA sequencing is prone to errors and therefore there is a need for dealing with the erroneous oligos produced by the sequencer. Church et al. in \cite{b1} has studied the main factors causing errors in the sequenced oligos. According to Church there are three main restrictions that should be respected:

\begin{itemize}
	\item \textbf{Homopolymers:} Consecutive occurencies of the same nucleotides should be avoided.

	\item \textbf{G, C content:} The percentage of G and C in the oligos should be lower or equal to the one of A and T.
	
	\item \textbf{Pattern repetitions:} The codewords used to encode the oligos should not be repeated forming the same pattern throughout the oligo length.
\end{itemize}
Taking into account all the above rules the sequencing error can be reduced. Consequently, in this work we  propose a novel efficient encoding algorithm which encodes the quantized wavelet coefficients using codewords that respect those biological constraints.
\\

\vspace{-3mm}
\subsection{Creating the codewords}
\label{ssec:codewords}
The encoding algorithm proposed in this paper takes into consideration all of the encoding restrictions described in \ref{ssec:restrictions}. Let $Q(x)=f(\alpha(x))$ be the quantized values $\hat{x}^i \in \Sigma$ produced by the quantizer with $i \in \{1,...,k\}$. $f$ is called the decoding function and $\alpha(x)=i$, $i \in \{1,...,k\}$, the encoding function providing the index of the quantization levels. In order to generate a DNA code $\Gamma$ we introduce two separate alphabets: 

\begin{itemize}
	\item $\mathcal{D}_1=\{AT,AC,AG,TA,TC,TG,CA,CT,GA,GT\}$
	\item $\mathcal{D}_2=\{A,T,C,G\}$
\end{itemize}

$\mathcal{D}_1$ is an alphabet composed by concatenations of two symbols from $\mathcal{D}_2$ selected in such a way that no homopolymers or high GC content is created. In order to encode the quantized sequence onto DNA we define the code $\Gamma$ as the application: $\Gamma:\Sigma\rightarrow \mathcal{D}^\star$ where $\mathcal{D}^\star$ is a dictionary composed by  $L \geq 2k$ codewords $c_i$ of length $l$. We denote $\Gamma(\hat{x}^i)=c_i$ the codeword associated with the quantized value $\hat{x}^i \in \Sigma$. $\mathcal{D}^\star$ is constructed by all the possible concatenations of symbols from $\mathcal{D}_1$ and $\mathcal{D}_2$ according to the following rules:

\begin{itemize}
	\item Codes with codewords of an even length ($l$ even) are being constructed selecting $\frac{l}{2}$ doublets from $\mathcal{D}_1$,
	\item Codes with codewords of an odd length ($l$ odd) are being constructed selecting $\frac{l-1}{2}$ doublets from $\mathcal{D}_1$ and a single symbol from $\mathcal{D}_2$.
\end{itemize}

\begin{figure}[h]
\centering
\includegraphics[width=0.5\columnwidth]{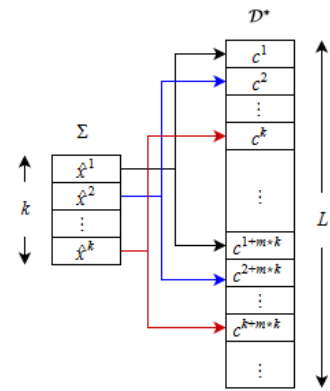}
\caption{Mapping the quantized values from codebook $\Sigma$ to codewords of ${\mathcal{D}^*}$ \vspace{-5mm}}
\label{fig:mapping}
\end{figure}

The quantization of wavelet subbands for big values of quantization step-size $q$ can lead to long repetitions of the same quantized value. The use of existing algorithms for the encoding of such a sequence into DNA would thus create pattern repetitions.
In this work, in  order to avoid those repetitions,
 we developed a new algorithm based on pseudorandom mapping which associates a quantized value to more than one possible codewords. More precisely our algorithm maps the index of levels of quantization $i$ to the codewords of $\mathcal{D ^*}$ as described in figure \ref{fig:mapping}. The code $\Gamma$ is constructed so that each quantized value in $\Sigma$ is mapped to a set of different non-empty quaternary codewords in $\mathcal{D}^\star$ following a one-to-many relation in such a way that it is uniquely decodable. Since we ensure $L \geq 2k$, the pseudorandom mapping can at least provides two possible codewords for one input symbol. More precisely, the mapping is described by the following steps:

\begin{enumerate}
\item Build the corresponding code $\mathcal{D^*}$ of size $L$ using all possible codewords of length $l$ which can be built following the two rules described previously,
\item Compute the number of times $m$ that $k$ can be replicated into the total size $L$ of the code $\mathcal{D^*}$: $m=\lfloor\frac{L}{k}\rfloor$, 
\item The mapping of the quantized value $\hat{x}_i$ to a codeword $c_i$ is given by: $\Gamma(\hat{x}^i)=\mathcal{D}^*(i+rand(0,m-1)*k )$.\\
\end{enumerate}

The encoding procedure is explained by Algorithm 1. It is obvious to prove that the code produced by this algorithm is uniquely decodable.

\vspace{0.25cm}
\begin{algorithm}[h]
\fontsize{9}{8}
\caption{Encoding Algorithm}
  \begin{algorithmic}[1]
    \State{Compute length $l$ of codewords needed for encoding all $k$ levels of quantization:}
   \IF{$\log_{10}{k}$ not an integer}
    \IF{$10^{\lfloor\log_{10}{k}\rfloor}*4\leq k$}
        \State$l=\lfloor\log_{10}{k}\rfloor*2+1$
        \ELSE{   $l=\lceil \log_{10}{k}\rceil*2$}
        \ENDIF
    \ELSE{$l=\log_{10}{k}*2$}
    \ENDIF
    \State{Build code $\mathcal{D}$ of $L$ different codewords:}
    \IF{$l$ is even}
    \State{Construct all possible codewords of length $l$ using $\frac{l}{2}$ choices from $\mathcal{D}_1$}
    \ELSIF{$l$ is odd}
    \State{Construct all possible codewords of length $l$ by using $\frac{l}{2}$ choices from 
     $\mathcal{D}_1$ adding one symbol from $\mathcal{D}_2$}
    \ENDIF
  \State{Mapping of index values of quantization to codewords from $\mathcal{D}$}
  \COMPUTE{$m=\lfloor\frac{L}{k}\rfloor$}
  \COMPUTE{$\Gamma(\hat{x}^i)=\mathcal{D^*}(i+rand(1,m-1)*k)$}
  \end{algorithmic}
\end{algorithm}
\vspace{-4mm}

\vspace{0.25cm}



\subsection{Formatting of the data}
\label{sec:formatting}
\begin{figure}[t]
\centering
\includegraphics[width=0.4\textwidth]{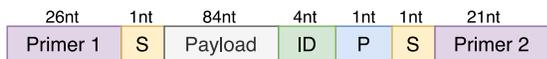}
\caption{Format of the oligos -  S denotes the sense nucleotide which determines whether a strand is reverse complemented when sequenced. P is a parity check nucleotide while the ID is an identifier of the image so to be distinguished from other data that may be stored. Payload contains encoded chunks, including specific headers. \vspace{-3mm}}
\label{format}
\end{figure}

As mentioned in \ref{ssec:general_idea}, the DNA synthesis is a biological procedure introducing a very small amount of error when oligos have a size that doesn't exceed 150 nts. However, this error increases exponentially as the oligos to be synthesized get longer. Consequently, the global encoded sequence needs to be cut into short chunks for generating the oligos. This implies the need for including headers inside the oligos, which contain information about the position of each of the chunks of encoded information, to allow the decoding and image reconstruction. Furthermore, the sequencing machines need some special sequences called primers that must be inserted both in the beginning and in the end of each oligo. Figure \ref{format} shows the way oligos are formatted before synthesis.

After the construction of the desired oligos the sequences are sent to be  chemically synthesized into synthetic DNA. The synthesized oligos are then being inserted into special capsules to be safely stored for many years.
\\
\vspace{-2mm}
\section{Data Decoding}
\label{sec:decoding}
\subsection{DNA Sequencing using bridge amplification (BA)}
\label{ssec:pcr}
In the decoding part of this encoding scheme, the data that has been encoded and synthesized into DNA must be retrieved by reading the synthetic oligos that have been stored into the storage capsule. This process is performed by special machines called sequencers. However as mentioned in \ref{ssec:general_idea}, sequencing is an error prone procedure and can cause significant errors as insertion, deletion or substitution of nucleotides. The use of a robust encoding scheme, like the one proposed by this work, which takes into consideration the biological restrictions that have been analytically described in section \ref{ssec:restrictions}, may reduce the probability of error occurences. However, in order to ensure the accuracy of the decoded data, the stored oligos are firstly cloned using PCR. Then the copies can be read by a sequencer using the method of BA which allows reading the oligos while cloning them into more copies. This redundancy is very important for the reduction of the sequencing error. One can imagine the role of the PCR and BA like the one of classical repetition coding used for transmission over a noisy channel that may corrupt the transmission in various positions. As in repetition coding, PCR and BA produces during sequencing many copies of the oligos hoping that only a minority of these copies will be corrupted by the sequencing noise.

\subsection{Selecting the non corrupted oligos}
\label{ssec:oligo_selection}
 As mentioned in section \ref{ssec:pcr} the sequencing method provides a great amount of copies, many of which are distorted with insertions deletions or substitutions.
 As many copies of oligos are corrupted by noise, it means that before reconstructing the image, the most representative copies of each oligo should be selected.
Consequently, a preprocessing procedure of data cleaning can importantly improve the quality of the sequenced data set of oligos provided by the sequencer. In our experiments we discarded all the oligos that did not match the exact length of 91 nucleotides or contained non-recognized bases 
Then, by checking the number of occurences of each oligo in the data set that matches the correct oligos size, we reconstruct each encoded subband using the most frequent oligos for each different chunks. This selection is based on the assumption that higher frequencies reveal a higher accuracy.

\vspace{-3mm}
\subsection{Image decoding}
After selecting the oligos that are not corrupted by the sequencing noise, one must decode the DNA sequences contained in each chunk (or equivalently in each oligo). As shown in section \ref{ssec:codewords}, since the proposed code is invertible it is possible to reconstruct the quantized subbands without any ambiguity thanks to the inverse mapping $\Gamma^{*}$. Then, by reassemblying the decoded chunks in each subband we get the decoded image in the wavelet domain. 
Finally, applying the inverse DWT we get a reconstruction of the initial input image.

\begin{table*}[t]
\resizebox{\linewidth}{!}{%
\begin{tabular}{|l|l|l|l|l|l|l|l|}
\hline
Parameter                  & Church et al.\cite{b2} & Goldman et al.\cite{b3}   & Grass et al.\cite{b4}  &Bornholt et al.\cite{b5}  & Blawat et. al \cite{b6}   & Erlich et al.\cite{b7} & Our work \\ \hline
Input data (Mbytes)        & 0.65          & 0.75       & 0.08  & 0.15       & 22        & 2.15     & 0.26    \\ \hline
Coding potential (bits/nt) & 1             & 1.58       & 1.78  & 1.58       & 1.6       & 1.98     & 2.14     \\ \hline
Redundancy                 & 1             & 4          & 1     & 1.5        & 1.13      & 1.07     & 1        \\ \hline
Error correction           & No            & Yes        & Yes   & No         & Yes       & Yes      & No       \\ \hline
Full recovery              & No            & No         & Yes   & No         & Yes       & Yes      & Yes      \\ \hline
Net information density    & 0.83          & 0.33       & 1.14  & 0.88       & 0.92      & 1.57     &  1.71    \\ \hline
Number of oligos           & 54,898        & 153,335    & 4,991 & 151,000    & 1,000,000 & 72,000   & 13,426      \\ \hline
\end{tabular}
}
\caption{Comparison to previous works - Coding potential: maximal information content of each nucleotide before indexing or error correcting. Redundancy: excess of synthesized oligos to provide robustness to dropouts. Error correction/ detection: the presence of error-correction code to handle synthesis and sequencing errors. Full recovery: DNA code was recovered without any error. Net information density: input information in bits divided by the number of synthesized DNA nucleotides (excluding primers).\vspace{-3mm}} 
\label{table:comparison}

\end{table*}

\vspace{-1mm}
\section{Results}
\label{sec:results}
\vspace{-1mm}
\subsection{Coding performance}
\begin{figure} [b]
\centering
\vspace{-1mm}
\includegraphics*[width=0.5\textwidth]{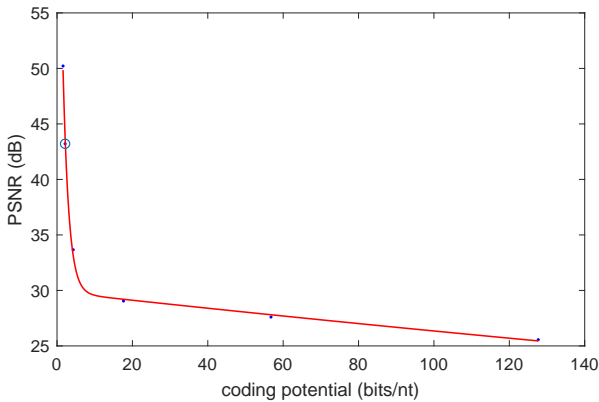}
\caption{ PSNR in function of the coding rate in bits per nucleotide for the image Lena of size 512 by 512 pixels. The selected point is the one represented in our results in table \ref{table:comparison}\vspace{-5mm}}
\label{PSNR-Rate}
\end{figure}

In our experiment, for the image compression we used a 3-level 9/7 DWT decomposition quantizing each subband using a uniform scalar quantizer with a quantization step size determined by an optimal bit-allocation algorithm.
 The quantized subbands have been then encoded using the proposed algorithm described in section \ref{ssec:codewords} respecting the restrictions imposed by the biological procedures. At this point it is important to highlight the fact that previous encoding procedures do not take into consideration the last restriction of pattern repetitions. In our case however, as we wish to use quantization of wavelet subbands in order to achieve high compression and decrease the synthesis cost, it is possible that after quantization we get a long sequence of repeated values in the quantized coefficients. The DNA coding of such a sequence can produce pattern repetitions which is an ill case as it is more likely for the sequencing to introduce more errors. Thus, our encoding algorithm tackles this problem by applying some kind of randomness as described in section \ref{ssec:codewords}. For the evaluation of our encoder's efficiency we used some compression measures that have been used also in \cite{b7}.\\
 Simulations have been carried out on Lena image of size 512 by 512 pixels. One can see on figure \ref{PSNR-Rate} the evolution of the Peak SNR (PSNR) in function of the coding rate in bits per nucleotide. In Table \ref{table:comparison} is reported the coding result at 2.14 bits per nucleotide which corresponds to a nearly lossless compression, allowing a fair comparison with the state-of-the art approaches. The corresponding PSNR is equal to $43.21 dB$ providing a perfect reconstructed image without any visual artefacts.

\subsection{Real oligonucleotides synthesis/sequencing}
\begin{figure}[t]
\centering
		\begin{minipage}[b]{0.5\linewidth}
				\centerline{
\includegraphics*[scale=0.9]{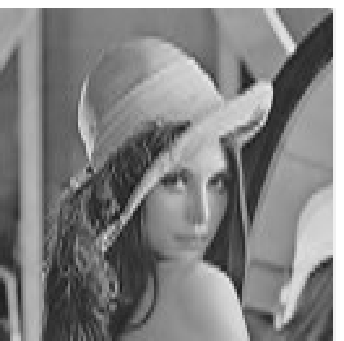}}
    \centerline{
		\begin{tabular}{c}
            \small PSNR=32.5 dB	
		\end{tabular}}
	   \end{minipage}
\hfill
		\begin{minipage}[b]{0.45\linewidth}
				\centerline{
\includegraphics*[scale=0.9]{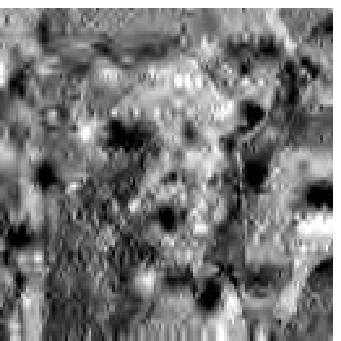}}
    \centerline{
		\begin{tabular}{c}
            \small PSNR=14.3 dB	
		\end{tabular}}
	   \end{minipage}\vspace{-2mm}

\caption{ Visual results for two different cases at 2.68 bits/nt: Reconstruction using the most frequent oligos (left) and random selection (right). For the left image the PSNR value is only due to the error inserted by the quantization process as we have managed to get a reconstruction without any sequencing noise.} \vspace{-4mm}
\label{fig:res}
\end{figure}

In this study we have carried out a real biological experiment for storing a small image of 128 by 128 pixels into DNA. The choice of the size of the image was constrained by the high expenses of the biological procedures involved in the experiment. Here, the coding leads to a PSNR equal to $32.5 dB$ at 2.68 bits per nucleotide. As shown in figure  \ref{fig:res} (left image), this experiment proves the feasibility of correctly retrieving back the stored image from DNA.
For the decoding part we have used the Illumina Next Seq sequencing machine. The sequencer provided us with a data set containing many copies of each stored oligo which also contained sequencing errors. In order to test the sequencer's reliability we tested two different ways of selecting the oligos for reconstructing the initial image. First we tested the reconstruction using the oligos with the highest frequencies which we assume to be the most representative ones. Then we also tested the case of random oligo selection. The visual results are presented in figure \ref{fig:res}. It is clear that by choosing the most frequent oligos from the different copies provided by the sequencing, it is more probable to achieve the best possible reconstruction of the image.

\vspace{-2mm}
\section{CONCLUSION}
\label{sec:conc}
In this paper we have introduced in the DNA storage workflow a new algorithm for the specific encoding of digital images into DNA, while also introducing for the first time and to our knowledge, image compression techniques in the process of DNA data storage. Compression allows to control easily the length of the generated DNA strands. The compression of the images to be stored into DNA is a very important improvement as this can reduce the synthesis cost. By this experiment we have managed to perfectly reconstruct the quantized input image while also providing very promising results in terms of achieved compression ratio. Those results motivate us to extend our work by  more robust encoding algorithms. Furthermore an interesting next step of this experiment would be the introduction of some error correction algorithm to treat the erroneous oligos and improve the efficiency of decoding.



%

%
%

\bibliography{references}






\end{document}